\documentclass[copyright]{eptcs}
\providecommand\event{MSFP 2018}

\usepackage{amsmath}
\usepackage{amssymb}
\usepackage{amsfonts}
\usepackage{mathrsfs}

\usepackage{amsthm}
\newtheorem{definition}{Definition}
\newtheorem{theorem}{Theorem}

\usepackage[strings]{underscore}
\usepackage{hyperref}

\newcommand\J{\mathscr J} 
\renewcommand\P{\mathscr P} 
\newcommand\R{\mathbb R} 
\newcommand\B{\mathbb B} 
\DeclareMathOperator*\argmax{\arg\max}

\title{Backward Induction for Repeated Games} \def\titlerunning{Backward Induction for Repeated Games}
\author{Jules Hedges \institute{University of Oxford}} \def\authorrunning{Jules Hedges}

\begin{document}

\maketitle

\begin{abstract}
	We present a method of backward induction for computing approximate subgame perfect Nash equilibria of infinitely repeated games with discounted payoffs.
	This uses the selection monad transformer, combined with the searchable set monad viewed as a notion of `topologically compact' nondeterminism, and a simple model of computable real numbers.
	This is the first application of Escard\'o and Oliva's theory of higher-order sequential games to games of imperfect information, in which (as well as its mathematical elegance) lazy evaluation does nontrivial work for us compared with a traditional game-theoretic analysis.
	Since a full theoretical understanding of this method is lacking (and appears to be very hard), we consider this an `experimental' paper heavily inspired by theoretical ideas.
	We use the famous Iterated Prisoner's Dilemma as a worked example.
\end{abstract}

\section{Introduction}

We present a method of backward induction for infinitely repeated games.
Since this is impossible, we more precisely perform $\epsilon$-backward induction, that is to say, we compute plays in which each player's choice gives them an outcome $\epsilon$-close to their optimal outcome, where $\epsilon$ is a small error bound.
We do this by combining Escard\'o and Oliva's interpretation of the \emph{product of selection functions} as an unbounded generalisation of backward induction \cite{escardo12}, together with the nondeterministic generalisation in \cite{escardo15b} and a view of the searchable set monad \cite{escardo04} as a notion of `compact' nondeterminism.
Since a full theoretical understanding of this method is lacking (and appears to be very hard), we consider this an `experimental' paper heavily inspired by theoretical ideas.

Backward induction is an algorithm for computing equilibria of games with sequential structure, known as early as Zermelo \cite{schwalbe01}.
The essence of backward induction is \emph{counterfactual reasoning}.
Suppose two players sequentially choose from sets $X, Y$, with payoffs given by $q_1, q_2 : X \times Y \to \R$, where $q_i (x, y)$ is the payoff of player $i$ given the choices $(x, y)$.
Player 1 reasoning as follows: suppose she chose $x$, then player 2 would choose $y$ in order to maximise $q_2 (x, y)$.
Ignoring for now that there may be several such $y$ associated to each $x$, this defines a function $f : X \to Y$.
(Accounting for different possible functions leads to games with multiple equilibria.)
Player 1 then chooses $x$ in order to maximise $q_1 (x, f (x))$.

Selection functions and the selection monad, introduced by Escard\'o and Oliva in \cite{escardo10a} and based on earlier work of Escard\'o in constructive topology \cite{escardo04}, present backward induction in such a way that it falls out of general considerations in category theory.
Moreover this reveals the non-obvious fact, often directly contradicted in standard game theory texts, that backward induction can be generalised to \emph{unbounded} games which have arbitrarily long plays \cite{escardo12}.

Both Zermelo's original presentation and Escard\'o-Oliva's generalisation are defined only for \emph{games of perfect information}, in which all players are aware of all relevant information.
However modern presentations of backward induction are defined for arbitrary games in extensive form: the additional step is that whenever a `simultaneous' subgame (i.e. a nontrivial information set) is reached, it should be replaced by a Nash equilibrium computed by some other method (see for example \cite[proposition 9.B.3]{mas_collel_etal_microeconomic_theory}).

An important class of games that include both simultaneous and sequential aspects are the \emph{repeated games} \cite{mailath_samuelson_repeated_games_reputations}, including the famous \emph{iterated prisoner's dilemma}.
In these games, a simple simultaneous `stage game' is played repeatedly (usually infinitely), with players at each stage being able to observe the complete play of all previous stages but not the action of other players in the current stage, and total payoffs being given by a convergent infinite sum.
The single-stage prisoner's dilemma is a simple model of the breakdown of trust and cooperation: in the game's unique Nash equilibrium (mutually stable choice of strategies) both players defect (betray each other) and receive less payoff than they could by cooperating.
By an argument using backward induction, in any \emph{finite} repetition of the prisoner's dilemma, both players will also always defect.
However the \emph{infinitely} repeated prisoner's dilemma also has cooperative equilibria such as tit-for-tat and grim trigger strategies, in which cooperation is enforced by the threat of later retribution.
(These games and strategies will all be defined later.)

In this paper we demonstration the application of Escard\'o and Oliva's theory to games of imperfect information, using the iterated prisoner's dilemma as a worked example.
We do this using the nondeterministic variant of the selection monad introduced in \cite{escardo15b}, defining a new operator called the \emph{sum of selection functions} that nondeterministically chooses a Nash equilibrium of a finite simultaneous game by brute force search, based on \cite{hedges_etal_selection_equilibria_higher_order_games}.
Two further innovations are required:
\begin{itemize}
	\item Defining the infinite discounted sum of payoffs requires using a simple model of computable reals, rather than an approximation such as rationals or machine-precision doubles.
	(Using an approximation would amount to disregarding play after finitely many stages, which drastically changes the game, for example ruling out cooperation in the iterated prisoner's dilemma.)
	Specialising to the iterated prisoner's dilemma and using Haskell allows us to essentially trivialise this step using lazy streams of digits in some base.
	However, the fact that computable reals do not have decidable inequality is a major crux, and prevents us from doing the impossible.
	Because of this restriction, we compute only \emph{approximate} equilibria.
	
	\item If we represent the nondeterministic choice of Nash equilibrium using representations based on the list or continuation monads, we find empirically that the resulting algorithm fails to terminate when combined with unbounded backward induction.
	Instead we use \emph{searchable sets}, which support unbounded backtracking.
	Viewing the searchable set monad as a notion of `compact' nondeterminism is a small but novel change of perspective.
\end{itemize}

This paper represents work done by the author several years ago as a continuation of \cite{hedges14}.
A good understanding of this method, and especially a game-theoretic understanding of the nondeterministic selection monad defined in \cite{escardo15b}, remains elusive.
It should therefore be considered as an `experimental' paper, presenting the method with intuitive justification but little theoretical analysis, which is currently ongoing work by the author with Joe Bolt and Philipp Zahn.

This paper switches between mathematical notation and Haskell.
A full Haskell implementation can be found at \url{http://www.cs.ox.ac.uk/people/julian.hedges/code/RepeatedGames.hs}.

\paragraph{Overview} In sections \ref{simultaneous} and \ref{sequential} we introduce simultaneous, sequential and repeated games.
In section \ref{selection} we define selection functions and implement them in Haskell.
In section \ref{product} and \ref{sum} we relate selection functions to sequential and simultaneous games respectively.
Sections \ref{searchable} and \ref{nondeterminism} concern searchable sets, and section \ref{reals} concerns computable reals.
Finally section \ref{together} puts all the pieces together, discusses the results and summarises problems for future work.

\section{Simultaneous games}\label{simultaneous}

In this section and the next we will introduce enough game theory from scratch in order to understand our worked example, the iterated prisoner's dilemma (IPD).

Informally, a \emph{game} in the sense of game theory is determined by the following data:
\begin{enumerate}
	\item A set of \emph{players}
	
	\item For each player, a set of \emph{choices} available at one or more points in the game at which that player makes a choice
	
	\item For each of those choices, a determination of the \emph{information} available when the choice is made; more precisely, a set of \emph{observations} that the player could make before making the choice; this determines a set of \emph{strategies} for the move, which are functions from observations to choices
	
	\item The `internal dynamics' of the game, which means a determination of the observation that will be made by each player given the choices of previous players; this requires that the temporal sequence of choices made is well-ordered
	
	\item For each player, a real number called the \emph{payoff}, determined by all choices made by all players (the \emph{play})
\end{enumerate}

This is traditionally formalised using \emph{extensive form games}, due to \cite{vonneumann44}.
However, since we only need certain special cases in this paper we will instead make several more specific definitions guided by this general template.
The extensive form represents a game via its \emph{tree} of plays, with choices taking place at nodes and payoff awarded at leaves; the information available to players is represented by a partition of nodes into `information sets' which cut across the tree structure.
More information can be found in any game theory textbook, for example \cite{mas_collel_etal_microeconomic_theory,brown08}.

\begin{definition}
	An $n$-player \emph{normal form game} is defined by sets of choices $X_i$ and payoff functions
	\[ q_i : \prod_{j = 1}^n X_j \to \R \]
	for each player $1 \leq i \leq n$.
\end{definition}

In a normal form, each player $i$ simultaneously makes a single choice from $X_i$.
The term `simultaneous' means that each player has no information available when making their choice.
This means that the set of possible observations is a singleton $\{ * \}$ where $*$ is a dummy observation representing `nothing observed', and so the strategy for player $i$'s (single) choice is nothing but an element of $X_i$, the choice itself.
(This is sometimes called a \emph{pure strategy}, to distinguish it from a \emph{mixed strategy} which is a probability distribution over choices.)

\begin{definition}
	A \emph{Nash equilibrium} of a normal form game is a tuple of choices $x : \prod_{i = 1}^n X_i$ with the property that for each player $1 \leq i \leq n$,
	\[ x_i \in \argmax_{x'_i : X_i} q_i (x [i \mapsto x'_i]) \]
	where $x [i \mapsto x'_i]$ is the tuple defined by
	\[ (x [i \mapsto x'_i])_j = \begin{cases}
		x'_i &\text{ if } i = j \\
		x_j &\text{ otherwise}
	\end{cases} \]
\end{definition}

The $\argmax$ operator is defined by
\[ \argmax_{x : X} k (x) = \{ x : X \mid k (x) \geq k (x') \text{ for all } x' : X \} \]
for $k : X \to \R$.

Informally, a Nash equilibrium is a tuple of strategies (called a \emph{strategy profile}) with the property that no player can strictly increase their payoff by deviating \emph{unilaterally} to a different strategy.
In this sense a Nash equilibrium is a combination of strategies that is self-enforcing, or more exactly, non-self-defeating.
A normal form game may have zero, one or many Nash equilibria.


Perhaps the most famous example of a normal form game is the \emph{prisoner's dilemma}.
This is a 2-player game in which the set of choices of the two players are $X_1 = X_2 = \{ C, D \}$, where $C$ stands for \emph{cooperate} and $D$ stands for \emph{defect}.
The payoffs are given as follows:
\[ q_1 (C, C) = 2 \qquad q_1 (C, D) = 0 \qquad q_1 (D, C) = 3 \qquad q_1 (D, D) = 1 \]
\[ q_2 (C, C) = 2 \qquad q_2 (C, D) = 3 \qquad q_2 (D, C) = 0 \qquad q_2 (D, D) = 1 \]
This game has exactly one Nash equilibrium, namely $(D, D)$.
Both players would receive a higher payoff if $(C, C)$ was played, but neither can trust the other to not `betray' them by deviating back to $D$.
In this sense, the prisoner's dilemma is a simple mathematical model of a breakdown of trust or cooperation.

%

\section{Sequential and repeated games}\label{sequential}

The next class of games we consider are the \emph{sequential games}.
These are a simplification due to Escard\'o and Oliva of the \emph{games of perfect information}, those games in which players can observe everything that happened in the past.
They come in two variants: \emph{bounded} and \emph{unbounded}.
We will present only the unbounded version, which is more general.

In a sequential game, the set of players is totally ordered, giving the order in which moves are made, and each player can observe the list of moves made by previous players.
(In a general game of perfect information, the player making a choice and the set of choices available may depend on the previous choices.)
We restrict to infinite games, in which the set of players is countably infinite, and the order-type is $\omega$.
(In particular, the game has a \emph{first} player who observes nothing.)
We further restrict to \emph{monomorphic} games, in which all players make choices from the same set.

\begin{definition}
	An unbounded monomorphic \emph{sequential game} is determined by a set $X$ of choices, together with a continuous function
	\[ q : X^\omega \to \R^\omega \]
	
	A \emph{strategy profile} of such a game is a function $\sigma : X^* \to X$ which, given a finite list $(x_1, \ldots, x_{i - 1})$ of choices observed by player $i$, gives the move $\sigma (x_1, \ldots, x_{i - 1})$ of player $i$.
\end{definition}

In the mathematical parts of this paper we carefully distinguish \emph{finite} lists $X^*$ from \emph{streams} $X^\omega$, although they are conflated in the Haskell code.
We will not formalise the meaning of \emph{continuous} in this paper, which requires some topology, but it roughly means that computing the output $q (x)$ to finite precision requires only knowing a finite prefix of $x$.
Games with a discontinuous payoff function, such as the dollar auction, require other techniques \cite{lescanne12} and can have more pathological behaviour.

Clearly, a strategy profile determines a stream of choices called its \emph{strategic play}.
More generally, given a strategy profile $\sigma : X^* \to X$ and a partial play $x = (x_1, \ldots, x_{i - 1}) : X^*$, we define a play $\nu^\sigma_x : X^\omega$, called the \emph{strategic extension} of $x$ by $\sigma$, by the course-of-values recursion
\[ (\nu^\sigma_x)_j = \begin{cases}
	x_j &\text{ if } j < i \\
	\sigma ((\nu^\sigma_x)_1^{j - 1}) &\text{ otherwise}
\end{cases} \]
This is the play which begins with $x$, and afterwards is played according to the strategies $\sigma$.

\begin{definition}
	A \emph{subgame perfect equilibrium} of a sequential game is a strategy profile $\sigma$ such that for all partial plays $x : X^*$ of length $i - 1$,
	\[ \sigma (x) \in \argmax_{x' : X} (q (\nu^\sigma_{(x, x')}))_i \]
	where $(x, x') : X^*$ is the sequence obtained by extending $x : X^*$ with $x' : X$.
\end{definition}

Next we turn our attention to \emph{repeated games}, which have both a sequential and a simultaneous aspect.
A repeated game comes from taking a simultaneous game and playing it infinitely often, summing the resulting payoffs.
A good introduction to repeated games can be found in chapter 2 of \cite{mailath_samuelson_repeated_games_reputations}.

A repeated game consists of a normal form \emph{stage game} played infinitely often, where in each stage the players can observe the choices made by all players in all previous stages, but not the choices of other places in the current stage.
If a player receives the infinite stream of payoffs $u : \R^\omega$ from the stage games, their total payoff is defined to be the \emph{discounted sum}
\[ u = \sum_{i = 1}^\infty \beta^i u_i \]
where $0 < \beta < 1$ is the \emph{discount factor}.
There are several ways to interpret the meaning of $\beta$.
It can be seen as a mere mathematical trick to make the total payoff converge, allowing us to avoid specifying explicit preferences on streams of payoffs.
It can be viewed as a measure of the `impatience' of the players, how much they prefer immediate utility to deferred utility, and alternatively $1 - \beta$ can be viewed as the probability that the game terminates after each round, with the discounted sum representing the \emph{expected} payoff.

\begin{definition}
	Given an $n$-player stage game with move sets $X_i$ and payoff functions $q_i : \prod_{j = 1}^n X_j \to \R$:
	\begin{itemize}
		\item A set of plays of the repeated game is $\left( \prod_{i = 1}^n X_i \right)^\omega$
		\item The $i$th player's payoff function $q_i^\infty : \left( \prod_{i = 1}^n X_i \right)^\omega \to \R$ is given by
		\[ q_i^\infty (x) = \sum_{k = 0}^\infty \beta^k q_i (x_k) \]
		where $0 < \beta < 1$ is a fixed discount factor
		\item A strategy for player $i$ in the resulting repeated game is a function
		\[ \sigma_i : \left( \prod_{j = 1}^n X_j \right)^* \to X_i \]
	\end{itemize}
\end{definition}

As for sequential games, a choice of strategy for each player determines a strategic play.
We also define a strategic extension operator for repeated games: Given a strategy profile $\sigma$ and a partial play $x : (\prod_{i = 1}^n X_i)^*$ with $j - 1$ stages, the stream $\nu^\sigma_x : (\prod_{i = 1}^n X_i)^\omega$ is defined by
\[ ((\nu^\sigma_x)_k)_i = \begin{cases}
	(x_k)_i &\text{ if } k < j \\
	\sigma_i ((\nu^\sigma_x)_1^{k - 1}) &\text{ otherwise}
\end{cases} \]
A strategy profile is called a \emph{subgame perfect equilibrium} if for all partial plays $x : \left( \prod_{i = 1}^n X_i \right)^*$ of length $j - 1$, and all players $1 \leq i \leq n$,
\[ \sigma_i (x) \in \argmax_{x'_i : X_i} q_i^\infty \left( \nu^\sigma_{\left( x, (\nu^\sigma_x)_j [i \mapsto x'_i] \right)} \right) \]
Here $\nu^\sigma_{\left( x, (\nu^\sigma_x)_j [i \mapsto x'_i] \right)} : \left( \prod_{i = 1}^n X_i \right)^\omega$ is the play where:
\begin{itemize}
	\item In the first $j - 1$ rounds, $x$ is played
	\item In the $j$th round, player $i$ plays $x'_i$ and all other players play according to $\sigma$
	\item In rounds greater than $j$, all players play according to $\sigma$
\end{itemize}

Consider a repeated form of the prisoner's dilemma, with $\beta = \frac{1}{4}$.
Given a stream of plays, the payoffs are given respectively by
\[ q^\infty_1 (x) = \sum_{i = 1}^\infty \frac{q_1 (x_i)}{4^i} \qquad\qquad q^\infty_2 (x) = \sum_{i = 1}^\infty \frac{q_2 (x_i)}{4^i} \]
where $q_1, q_2$ are the payoff functions given in the previous section.
One example of a subgame perfect equilibrium is to play the stage equilibrium $(D, D)$ in every stage irrespective of earlier play.
If we only \emph{finitely} repeat the prisoner's dilemma (equivalent to modifying $q^\infty_1, q^\infty_2$ to use only a finite sum), this is the only subgame perfect equilibrium.
However, in the infinitely repeated game there are many subgame perfect equilibria, some of which have plays in which $C$ is always played.
Possible payoffs resulting from subgame perfect equilibria of infinitely repeated games are characterised by the \emph{folk theorems} \cite[chapter 3]{mailath_samuelson_repeated_games_reputations}.

An example of a cooperative subgame perfect equilibrium is as follows.
Player 1 plays the strategy \emph{tit for tat}:
\[ \sigma_1 (x_1, \ldots, x_{j - 1}) = \begin{cases}
	C &\text{ if } j = 1 \\
	(x_{j - 1})_2 &\text{ otherwise}
\end{cases} \]
which initially cooperates, and otherwise copies the opponent's previous move.
Player 2 plays the strategy \emph{grim trigger}:
\[ \sigma_2 (x_1, \ldots, x_{j - 1}) = \begin{cases}
	C &\text{ if } (x_k)_1 = C \text{ for all } k < j \\
	D &\text{ otherwise}
\end{cases} \]
which cooperates as long as the opponent cooperates, but defects for all time if the opponent defects.
The strategic play of this subgame perfect equilibrium is $(C, C)$ in every stage.

\section{Selection functions}\label{selection}

In this section and the next we recall the theory of selection functions, which was developed mostly by Escard\'o and Oliva and can be found in many references including \cite{escardo04,escardo10a,escardo10d,escardo11,escardo15b}.

A \emph{selection function} is a function of type $(X \to R) \to X$.
We write this type as $\J_R (X)$.
More generally, given a type constructor $T$, a $T$-selection function is a function of type $(X \to R) \to T (X)$, which we write as $\J^T_R (X)$.
In Haskell:
\begin{verbatim}
	newtype SelT r t x = SelT {runSelT :: (x -> r) -> t x}
	
	instance (Functor t) => Functor (SelT r t) where
	  fmap f (SelT e) = SelT (\k -> fmap f (e (k . f)))
\end{verbatim}

For example, working for a moment in set theory, the $\argmax$ operator over a set $X$ is a selection function of type $\argmax : \J^\P_\R (X)$, where $\P$ is powerset and $\R$ is the set of real numbers.
This operator takes a function $k : X \to \R$ to the set
\[ \argmax (k) = \argmax_{x : X} k (x) = \{ x \in X \mid k (x) \geq k (x') \text{ for all } x' \in X \} \]
If $X$ is nonempty and finite then $\argmax (k)$ is nonempty and finite for every $k : X \to \R$.
Another example of a $\P$-selection function is $\textrm{fix} : \J^\P_X (X)$, which takes every function $k : X \to X$ to the set $\textrm{fix} (k) = \{ x \in X \mid x = k (x) \}$.

If $T$ is a strong monad and $R$ is a $T$-algebra then $\J^T_R$ can be given the structure of a strong monad \cite{escardo15b}.
(Since every type is an algebra of the identity monad, $\J_R = \J^{\operatorname{Id}}_R$ is a strong monad for every $R$.)
We begin by setting up a Haskell typeclass for $T$-algebras (requiring the \texttt{MultiParamTypeClasses} and \texttt{FlexibleInstances} language extensions):
\begin{verbatim}
	class Algebra t a where structure :: t a -> a
	instance Algebra Identity a where structure = runIdentity
	instance (Functor t, Algebra t x, Algebra t y) => Algebra t (x, y) where
	  structure a = (structure (fmap fst a), structure (fmap snd a))
\end{verbatim}

The definition of the monad structure on $\J_R^T$ is as follows:
\begin{verbatim}
	instance (Monad t, Algebra t r) => Monad (SelT r t) where
	  return = SelT . const . return
	  SelT e >>= f = SelT (\k -> let g x = runSelT (f x) k
	                                 h x = structure (fmap k (g x))
	                              in e h >>= g)
\end{verbatim}
This is admittedly a hard definition to understand.
A relatively gentle explanation, relating $\J_R^T$ to the continuation monad, can be found in \cite{hedges15a}.


In order to compute $\argmax$ with outcomes in an ordered type, we need to do a brute force search.
To do this cleanly we define a type class for \emph{finite} types, which have an exhaustive list of elements:
\begin{verbatim}
	class (Eq x) => Finite x where exhaust :: [x]
	instance (Finite x, Finite y) => Finite (x, y) where
	  exhaust = [(x, y) | x <- exhaust, y <- exhaust]
\end{verbatim}

Now we can define:
\begin{verbatim}
	argmax :: (Finite x, Ord r) => SelT r [] x
	argmax = SelT (\k -> [x | x <- exhaust, all (\x' -> k x >= k x') exhaust])
\end{verbatim}

A useful fact about the selection monad is that it is contravariant in the outcome type: Given a function $f : S \to R$ we obtain a monad morphism $f^* : \J^T_R \to \J^T_S$ \cite[section 1.1.8]{hedges_towards_compositional_game_theory}.
In Haskell:
\begin{verbatim}
	reindex :: (s -> r) -> SelT r t x -> SelT s t x
	reindex f (SelT e) = SelT (\k -> e (f . k))
\end{verbatim}
(This should be contrasted with the continuation monad, which is not functorial in the outcome type.)
In particular, reindexing $\argmax : \J^\P_\R (X)$ by the $i$th projection $\pi_i : \R^n \to \R$ yields the selection function $\pi_i^* (\argmax) : \J^\P_{\R^n} (X)$ of the $i$th player in an $n$-player game, who optimises the $i$th coordinate of the outcome and is indifferent about the others.

\section{The product of selection functions}\label{product}

Let $T$ be a strong monad and $R$ a $T$-algebra.
As a strong monad, $\J^T_R$ admits a binary \emph{monoidal product}
\[ \otimes : \J^T_R (X) \times \J^T_R (Y) \to \J^T_R (X \times Y) \]
In Haskell's do-notation, this monoidal product operator is especially intuitive:
\begin{verbatim}
	otimes :: (Monad t) => t x -> t y -> t (x, y)
	otimes a b = do {x <- a; y <- b; return (x, y)}
\end{verbatim}
We can also fold this operator across finite lists to give $\bigotimes : \J^T_R (X)^* \to \J^T_R (X^*)$, and across streams to give $\bigotimes : \J^T_R (X)^\omega \to \J^T_R (X^\omega)$.
Both of these folds are implemented by the Haskell prelude function \texttt{sequence :: (Monad t) => [t a] -> t [a]}; we will return to the question of productiveness on streams (i.e. whether each element is computed in finite time) later.

When $T$ is the identity monad, the following fundamental theorem connects selection functions with game theory \cite{escardo10a,escardo11,escardo12}:
\begin{theorem}
	Let $\mathcal G$ be a monomorphic unbounded sequential game defined by the choice set $X$ and the continuous outcome function
	\[ q : X^\omega \to \R^\omega \]
	For each $i \geq 1$ let $\epsilon_i : \J_{\R^\omega} (X)$ be a selection function such that $\epsilon_i (k) \in \argmax (\pi_i \circ k)$ for every $k : X \to \R^\omega$.
	Then
	\[ \left( \bigotimes_{i = 1}^\infty \epsilon_i \right) (q) : X^\omega \]
	is well-defined and is the strategic play of a subgame perfect equilibrium of $\mathcal G$.
\end{theorem}

In fact, this theorem has nothing to do with the $\argmax$ operator: Escard\'o and Oliva define \emph{higher order} sequential games whose definition involves selection functions, and prove that the product of selection functions computes plays of subgame perfect equilibria in this more general case.

This infinite product can be directly implemented in Haskell using \texttt{sequence}, producing a productive stream giving the play of a subgame perfect equilibrium \cite{escardo10d}.

We briefly digress to consider the (largely not understood) game-theoretic meaning of the monad $\J^T_R$ where $T$ is the nonempty finite powerset monad.
The monoidal product of this monad is used for proof-theoretic purposes in \cite{escardo15b}.
For simplicity we consider a \emph{finite} sequential game with $n$ stages, given by the payoff function $q : X^n \to \R^n$.

For nonempty finite $X$, $\pi_i^* (\argmax)$ itself has the type $\J^T_{\R^n} (X)$.
It is therefore reasonable to ask whether there is a choice of $T$-algebra (affine semilattice) structure on $\R^n$ such that the (nonempty finite) set
\[ \left( \bigotimes_{i = 1}^n \pi_i^* (\argmax) \right) (q) : T (X^n) \]
is the set of \emph{all} strategic plays of subgame perfect equilibria.
This does not appear to be the case, however.
Characterising sets that can be defined this way in game-theoretic terms is ongoing work with Joe Bolt and Philipp Zahn.

\section{The sum of selection functions}\label{sum}

Previous work on the product of selection functions (for example \cite{escardo10a,escardo10d,escardo11}) has considered only games of perfect information, which means that whenever a player makes a choice, they have access to all relevant information about the choices of other players.
More precisely, `having access' means that their strategy is a function that can depend on this information.
However, repeated games such as iterated prisoner's dilemma are not games of perfect information, but rather have both simultaneous and sequential aspects.
In this section we suggest a way to handle simultaneous choices in the selection function paradigm.

Games defined by explicit selection functions are called `higher order games', by analogy to selection functions being higher order functions.
In \cite{hedges_etal_selection_equilibria_higher_order_games} a solution concept suitable for simultaneous higher order games was considered, called \emph{selection equilibrium}.

\begin{definition}
	A 2-player \emph{higher order simultaneous game} consists of the following data:
	\begin{itemize}
		\item Sets $X$, $Y$ of \emph{choices} for the two players
		\item A set $R$ of \emph{outcomes} and an \emph{outcome function} $q : X \times Y \to R$
		\item For each player a \emph{multi-valued selection function} $\epsilon : \J^\P_R (X)$, $\delta : \J^\P_R (Y)$
	\end{itemize}
	A \emph{selection equilibrium} is a pair $(x, y) : X \times Y$ such that $x \in \epsilon (\lambda x' . q (x', y))$ and $y \in \delta (\lambda y' . q (x, y'))$.
\end{definition}

If $R = \mathbb R^2$, $\epsilon = \pi_1^* (\argmax)$ and $\delta = \pi_2^* (\argmax)$ then the definition of selection equilibrium reduces to ordinary Nash equilibrium.
Note that here we do \emph{not} assume that $R$ is a $\P$-algebra, so $\J^\P_R$ may not be a monad.

The definition of selection equilibria crucially relies on the selection functions being multi-valued.
However, the product of selection functions is studied mainly for single-valued selection functions, and its generalisation to nondeterministic selection functions is poorly understood in game-theoretic terms.
This barrier has prevented a unification of higher-order sequential and simultaneous games.

We propose the following definition, called the \emph{sum of selection functions}, which returns the set of selection equilibria, analogously to the product of selection functions for subgame perfect equilibria.
(The terminology `sum' is based on an optimistic hope that it has a nice algebraic interaction with the product of selection functions, although this is left for future work.)
\begin{definition}
	The binary operator
	\[ \oplus : \J^\P_R (X) \times \J^\P_R (Y) \to \J^\P_R (X \times Y) \]
	is defined by
	\[ (\epsilon \oplus \delta) (k) = \{ (x, y) : X \times Y \mid x \in \epsilon (\lambda x' . k (x', y)) \text{ and } y \in \delta (\lambda y' . k (x, y')) \} \]
\end{definition}

A problem with $\oplus$ is that $(\epsilon \oplus \delta) (q)$ can be empty even if $\epsilon (k)$ and $\delta (k)$ are never empty.
For example, if $X = Y = \{ A, B \}$ and $q : X \times Y \to \R$ is defined by
\[ q (x, y) = \begin{cases}
	1 &\text{ if } x = y \\
	0 &\text{ if } x \neq y
\end{cases} \]
then $(\argmax \oplus \arg\min) (q) = \varnothing$.
(This is the game \emph{matching pennies}.)
While it is often demanded that nondeterminism is represented by \emph{non-empty} powerset for partly philosophical reasons (the empty set represents \emph{failure} or partiality of a computation), in this paper we have a specific reason to be wary of it: In the next section we are going to represent nondeterminism using searchable sets, and the empty set is not searchable.

We will ignore this problem in this paper because it does not come up in the iterated prisoner's dilemma example; in the next section when we define the function \texttt{searchList} we will make it throw an exception on the empty list, and we find in practice that no exception is thrown.
(Dealing with this problem properly will require some more work, for example adding probabilistic strategies and relying on Nash's theorem \cite{nash50b}, or alternatively adding an explicit `failure' strategy using an exception monad, with explicit strategic preferences defined over the exception using monad algebras.)

We then define the sum of selection functions by a brute force search over finite sets.
\begin{verbatim}
	oplus :: (Finite x, Finite y) => SelT r [] x -> SelT r [] y -> SelT r [] (x, y)
	oplus (SelT e) (SelT d) = SelT (\k -> [(x, y) | (x, y) <- exhaust,
	  x `elem` (e (\x' -> k (x', y))),
	  y `elem` (d (\y' -> k (x, y')))])
\end{verbatim}

We now demonstrate that the sum of selection functions correctly computes the unique Nash equilibrium of the (single-stage) prisoner's dilemma.
We define our stage game, the prisoner's dilemma, in Haskell:
\begin{verbatim}
	data Move = C | D deriving (Show, Eq)
	instance Finite Move where exhaust = [C, D]
\end{verbatim}

Although we could define outcomes as integers or doubles, we will instead define a specific datatype, since we will be reusing it later as part of our representation of computable reals.
We call it `quit', short for \emph{quaternary digit}
\begin{verbatim}
	data Quit = Zero | One | Two | Three deriving (Show, Eq, Ord)
\end{verbatim}
Note that the \texttt{Ord} instance derived by Haskell is $0 < 1 < 2 < 3$.

Now the payoff function of prisoner's dilemma is
\begin{verbatim}	
	pd :: (Move, Move) -> (Quit, Quit)
	pd (C, C) = (Two, Two)
	pd (C, D) = (Zero, Three)
	pd (D, C) = (Three, Zero)
	pd (D, D) = (One, One)
\end{verbatim}

The first and second players respectively choose a single move to maximise the first and second coordinate of the outcome, given by \texttt{reindex fst argmax} and \texttt{reindex snd argmax}.
We take the sum of these selection functions, and apply it to the outcome function \texttt{pd}:
\begin{verbatim}
	> runSelT (reindex fst argmax `oplus` reindex snd argmax) pd
	==> [(D,D)]
\end{verbatim}
%
%
%

\section{Searchable sets}\label{searchable}

The property of the monad $\J_R$ that makes it suitable for working with unbounded games is that it supports an infinite \emph{monoidal product} $\bigotimes : \J_R (X)^\omega \to \J_R (X^\omega)$.
In particular, recall from section \ref{product} that the Haskell Prelude function \texttt{sequence :: (Monad m) => [m x] -> m [x]}, when specialised to the selection monad, is productive on infinite lists.

Monads for which \texttt{sequence} is productive on infinite lists in Haskell include the identity, state, \texttt{IO}, reader, writer and selection monads, and monad transformer stacks containing only these.
Monads for which \texttt{sequence} is not productive include the \texttt{Maybe}, list and continuation monads, and monad transformer stacks containing any of these.
These are empirical observations only: a theoretical characterisation of these monads is lacking. Such a characterisation would have to explain the large difference in behaviour between the seemingly similar types $(X \to R) \to X$ (selection monad) and $(X \to R) \to R$ (continuation monad).

In particular, the monad $\J_R^T$ does not admit infinite products when $T$ is the list monad, which is the most straightforward and common representation of nondeterminism in Haskell, and the one we used in section \ref{sum}.
That is to say, if we simply apply unbounded backward induction at this point then the resulting function will fail to terminate.
We must instead find a suitable alternative to the list monad to represent nondeterminism.
We find this in the \emph{searchable set monad}, which is the special case $\J_\B$ of the selection monad \cite{escardo04,escardo08}.

\begin{definition}
	A \emph{subset} $S \subseteq X$ of a type $X$ is defined to be its characteristic function $\chi_S : X \to \B$.
	For an element $x : X$, we write $x \in S$ if $\chi_S (x)$ is \emph{\texttt{True}}.
	A subset $S$ is called \emph{searchable} if there is a selection function $\epsilon : \J_\B (X)$ with the following two properties:
	\begin{itemize}
		\item $\epsilon (p) \in S$ for all predicates $p : X \to \B$
		\item For all $p : X \to \B$, if there exists an element of $S$ satisfying $p$ then $\epsilon (p)$ is such an element
	\end{itemize}
	If this is the case, we say that $\epsilon$ \emph{represents} $S$.
\end{definition}

This is a constructive analogue of topological \emph{compactness}.
Notice however that the empty subset is never searchable, since the selection function must always return an element of the subset.

We set up a Haskell type for this special case:
\begin{verbatim}
	type Searchable = SelT Bool Identity 
	
	searchable :: ((x -> Bool) -> x) -> Searchable x
	searchable e = SelT (Identity . e)
	
	search :: Searchable x -> (x -> Bool) -> x
	search (SelT e) = runIdentity . e
\end{verbatim}

Given any nonempty finite list of elements of a type $X$, we can produce a searchable set containing only those elements by searching the list:
\begin{verbatim}
	searchList :: [x] -> Searchable x
	searchList [] = error "searchList: Empty list"
	searchList xs = searchable (\p -> case Data.List.find p xs of
	  Nothing -> head xs
	  Just x  -> x)
\end{verbatim}
Using \texttt{searchList}, we can immediately `promote' a list-based nondeterministic selection function to a searchable set-based one:
\begin{verbatim}
	promote :: SelT r [] x -> SelT r Searchable x
	promote (SelT e) = SelT (searchList . e)
\end{verbatim}

Searchable sets are closed under several constructions analogous to compact topological spaces, notably forward images using \texttt{fmap} and countable products using \texttt{sequence}.
The latter is a constructive form of the countable Tychonoff theorem, and is used to produce `seemingly impossible functional programs' that search the Cantor space $2^\omega$ (or \texttt{[Bool]}) in finite time \cite{escardo07}.

Searchable sets admit decidable existential and universal quantification.
Given a selection function $\epsilon : \J_\B (X)$ for a searchable set $S \subseteq X$ and a predicate $p : X \to \B$, we know that if any element of $S$ satisfies $p$ then $\epsilon (p)$ is such an element.
Therefore in order to check whether any element of $S$ satisfies $p$ it suffices to test whether $\epsilon (p)$ satisfies $p$.
In Haskell:
\begin{verbatim}
	exists :: Searchable x -> (x -> Bool) -> Bool
	exists e p = p (search e p)
\end{verbatim}
The universal quantifier is defined by de Morgan duality $\forall = \neg \exists \neg$:
\begin{verbatim}
	forall :: Searchable x -> (x -> Bool) -> Bool
	forall e p = not (exists (not . p))
\end{verbatim}

\section{Compact nondeterminism}\label{nondeterminism}

It is worth saying a few words on the view of the searchable set monad as a `notion of nondeterminism'.
In Haskell it is possible to use the list monad to express a logic programming style of backtracking search.
(The use of the nonempty powerset monad to represent nondeterminism is due to \cite{moggi91}, based on earlier work on powerdomains in domain theory.)
For example, suppose we define a binary nondeterministic choice operator as follows:
\begin{verbatim}
	type Choice a = [a]

	choose :: a -> a -> [a]
	choose x y = [x, y]
\end{verbatim}
Now consider the following program:
\begin{verbatim}
	choose2 :: Choice (Int, Int)
	choose2 = do {x <- 0 `choose` 1; y <- 0 `choose` 1; return (x, y)}
\end{verbatim}
This program searches through the cartesian product $\{ 0, 1 \}^2$.
Given a predicate implemented as a Haskell function \texttt{p :: (Int, Int) -> Bool}, we can use a function like \texttt{Data.List.find} to `resolve' the nondeterminism to a deterministic search for a satisfying input of \texttt{p}.

The previous program can be rewritten using the function \texttt{sequence}, which performs a list of actions in a monad in order:
\begin{verbatim}
	choose2' :: Choice (Int, Int)
	choose2' = do {[x, y] <- sequence [0 `choose` 1, 0 `choose` 1]; return (x, y)}
\end{verbatim}
This form suggests attempting hypercomputation with an infinite sequence of nondeterministic choices searching for a satisfying input to a predicate \texttt{p :: [Int] -> Bool} on streams:
\begin{verbatim}
	chooseInfinity :: Choice [Int]
	chooseInfinity = sequence (repeat (0 `choose` 1))
\end{verbatim}
Mathematically, \texttt{sequence} for the list monad computes cartesian products, and so this program seems like it should enumerate the (set-theoretically uncountable) Cantor set $\{ 0, 1 \}^\omega$.
Unsurprisingly this fails and the program does not terminate, but Haskell is unable even to produce the obvious first element \texttt{[0,0,0,\dots]}:
\begin{verbatim}
	> sequence (repeat (0 `choose` 1))
	==> *** Exception: stack overflow
\end{verbatim}
(Intriguingly, the stack overflow happens \emph{before} the opening square bracket is printed.)

If we replace the list monad with the searchable set monad, however, we can do precisely this.
\begin{verbatim}
	chooseS :: a -> a -> Searchable a
	chooseS x y = searchList (choose x y)
	-- or chooseS x y = searchable (\p -> if p x then x else y)
	
	choose2'' :: Searchable (Int, Int)
	choose2'' = do {x <- 0 `chooseS` 1; y <- 0 `chooseS` 1; return (x, y)}
\end{verbatim}
Given a predicate \texttt{p :: Int -> Int -> Bool}, we resolve the nondeterministic choices to a deterministic search using \texttt{runSearchable choose2'' p}, which will return a satisfying input if one exists.
Furthermore, this now extends to infinite search:
\begin{verbatim}
	chooseInfinity' :: Searchable [Int]
	chooseInfinity' = sequence (repeat (0 `chooseS` 1))
\end{verbatim}
This can also be written more suggestively using direct recursion:
\begin{verbatim}
	chooseInfinity'' :: Searchable [Int]
	chooseInfinity'' = do x <- 0 `chooseS` 1
	                      xs <- chooseInfinity''
	                      return (x : xs)
\end{verbatim}
Now given some \texttt{p :: [Int] -> Bool}, the stream \texttt{runSearchable chooseInfinity' p} is productive and satisfies \texttt{p} if possible.
This is a \emph{seemingly impossible functional program}.

The reason that this is possible is that as a computable function, the output of \texttt{p} can only depend on a finite prefix of its input stream, and lazy evaluation only evaluates as much as is needed.
However, the operational behaviour of the seemingly impossible functional programs is still not well understood.

\section{Computable reals}\label{reals}

Our stage game, the prisoner's dilemma, has outcomes in $\{ 0, 1, 2, 3 \}$, and the iterated game is further parameterised by a discount factor $0 < \beta < 1$.
For convenience we choose the the discount factor to be $\beta = \frac{1}{4}$.
This means that the discounted sum
\[ \sum_{i = 1}^\infty \frac{a_i}{4^i} \]
can be represented as an infinite stream of digits in base-4, where the $i$th digit is precisely $a_i$.
That is to say, the discounted sum is represented by the \emph{identity function} on streams.

Recalling the type of quaternary digits we defined in section \ref{sum}, we define a Haskell type of real numbers in the unit interval represented as streams of quaternary digits, or quit-streams:
\begin{verbatim}
	data R = [Quit]
\end{verbatim}
Due to our trick of choosing a base-4 representation, we can avoid needing to define arithmetic operations on infinite quit-streams in the specific example of the iterated prisoner's dilemma.
This is fortunate, because they cannot be defined in general.
More generally, representations of real numbers based on infinite streams of digits (which are possibly the most naive or obvious representation, at least in a language such as Haskell with direct support for streams) support constructive \emph{topology}, but not constructive \emph{arithmetic}.
There does exist a model of the real numbers that supports both arithmetic and topology constructively: the \emph{signed digit stream} model \cite{weihrauch-simple-introduction-computable-analysis,escardo98}.
Using this would be necessary for a more general implementation.

The important piece of work we have to do is to define an approximate ordering on $\texttt{R}$.
As is well known, computable real numbers do not admit a computable ordering.
They do admit a semi-computable ordering that terminates except on a set of measure zero, namely to search along a pair of streams looking for a difference and simply run forever if the streams are equal.
However we find in practice that the product of selection functions does indeed search this diagonal, and will fail to terminate if we do this.

At this point we fix a positive integer $N$, written \texttt{precision} in Haskell, which is the number of significant quaternary digits.
In practice, we can only search to $N = 4$ in a reasonable amount of time, although $N = 5$ might be possible with enough optimising and a fast CPU.
(The runtime of the product of selection functions has not been theoretically characterised, but is likely to be very fast-growing and is possibly nonelementary, i.e. the runtime grows as a stack of exponentials of height $N$.)
\begin{verbatim}
	precision = 4 :: Int
\end{verbatim}

A fact about digit stream representations is that they admit non-identity equalities, exemplified by the (in)famous fact that $1 = 0.999\ldots$.
Similarly, in the lazy base 4 representation \texttt{One : repeat Zero} denotes the same real number as \texttt{Zero : repeat Three}.
This is a general fact about computable analysis (precisely, the fact is that the embedding $\texttt{R} \to [0, 1]$ of any model into the `true' unit interval necessarily fails to be injective).

When dealing with repeated games we should take particular care about this, because it is crucial for cooperative equilibria such as tit-for-tat being subgame perfect that a finite payoff now can be exactly balanced by future payoffs.
For example, `betraying' your opponent in order to receive a payoff of $1$ now followed by receiving $0$ forever after in eternal punishment, should be considered equally desirable as receiving $0$ now followed by an infinite reward of $3$ in every future stage.

Motivated by this, we compare approximate equality by evaluating the first $N$ digits of our quit streams as doubles, and then comparing to order $\frac{1}{4^{N - 1}}$.
This is reasonable since $N$ is small in practice.
\begin{verbatim}
	quit2Double :: Quit -> Double
	quit2Double x = case x of {Zero -> 0.0; One -> 1.0; Two -> 2.0; Three -> 3.0}

	real2Double :: R -> Double
	real2Double xs = sum (zipWith f xs [1 .. precision])
	  where f x n = quit2Double x * 0.25^n
	  
	greater :: R -> R -> Bool
	greater xs ys = real2Double xs > real2Double ys - 0.25^(precision - 1)
\end{verbatim}

\texttt{greater xs ys} computes \texttt{xs} $\geq$ \texttt{ys} to precision $N$, i.e. if \texttt{xs} is slightly smaller than \texttt{ys} but we must search further than $N$ digits to discover the fact, then \texttt{greater xs ys} returns \texttt{True}.
(Note that the Haskell function \texttt{zipWith}, when presented with lists of different lengths, will truncate the longer list.
Since \texttt{xs} is infinite, \texttt{real2Double} takes the first $N$ digits of it.)

Armed with the approximate ordering, we can now implement $\frac{1}{N}\text{-}\argmax$ as a nondeterministic selection function, using Haskell's lists monad as a basic representation of nondeterminism.
\begin{verbatim}
	eargmax :: (Finite a) => SelT Double [] a
	eargmax = SelT (\k -> [x | x <- exhaust, all (\x' -> k x `greater` k x') exhaust])
\end{verbatim}
Since every function (on a finite set) has an attained maximum, \texttt{eargmax} always returns a nonempty list.

Furthermore, using quantifiers for searchable sets we can define an algebra $\max : \J_\B (\R) \to \R$ which, given a selection function representing a searchable set $S \subseteq \R$, searches for an element of $x \in S$ satisfying $\forall y \in S . x \geq y$.
(Apart from the use of approximate inequality, this exhibits the elementary real analysis fact that a compact set of reals contains its maximum.)
In Haskell:
\begin{verbatim}
	instance Algebra Searchable R where
	  structure e = search e (\x -> forall e (\y -> x `greater` y))
\end{verbatim}

Unfortunately, the use of approximate inequality means that this does not obey the axioms of a monad algebra.
We proceed anyway since it appears to work in practice.
If the reader is worried abut this, we could equivalently make the type of outcomes $\J_\B (\R)$ (or \texttt{Searchable R}), which is the free $\J_\B$-algebra on $\R$, and move this use of the approximate ordering into the selection function (whose type becomes $(X \to \J_\B (\R)) \to \J_\B (X)$).
That is, instead of using the standard $\argmax$ function directly, we generalise it to a `nondeterministic $\argmax$' that imposes its own ordering on compact sets of outcomes.
Such variants of $\argmax$ are considered in \cite{hedges14}.

\section{Putting it together}\label{together}

The payoff function of the \emph{iterated} prisoner's dilemma takes a stream of pairs of choices, and computes the discounted sum of payoffs according to \texttt{pd}.
Due to our choice of representation, this is trivial:
\begin{verbatim}
	ipd :: [(Move, Move)] -> (R, R)
	ipd ms = (map (fst . pd) ms, map (snd . pd) ms)
\end{verbatim}

We demonstrate that applying the product of selection functions directly over the list monad does not terminate:
\begin{verbatim}
	stage :: SelT (R, R) [] (Move, Move)
	stage = reindex fst eargmax `oplus` reindex snd eargmax

	> :type runSelT (sequence (repeat stage)) ipd
	==> runSelT (sequence (repeat stage)) ipd :: [[(Move, Move)]]
	> runSelT (sequence (repeat stage)) ipd
	==> *** Exception: stack overflow
\end{verbatim}

If we first promote the stage game from the list to the searchable set monad, we obtain a searchable set of plays rather than a list of them:
\begin{verbatim}
	plays :: Searchable [(Move, Move)]
	plays = runSelT (sequence (repeat (promote stage))) ipd
\end{verbatim}
In order to obtain an element of this set, we must supply it with a predicate.
If we give the constant true predicate, we will obtain an arbitrary element of the set:
\begin{verbatim}
	> :type search plays (const True)
	==> search plays (const True) :: [(Move, Move)]
\end{verbatim}
Since this is an infinite list (a play of the repeated game) we request a finite prefix:
\begin{verbatim}
	> take 6 (search plays (const True))
	==> [(D,D),(D,D),(D,D),(C,C),(C,C),(C,C)]
\end{verbatim}
With precision set to 4, this takes around 5 minutes to run (interpreted) on the author's laptop.

With a little experimentation, we find that the searchable set of plays contains precisely the plays whose first three stages are \texttt{(D,D)}, i.e. the plays that have \texttt{[(D,D),(D,D),(D,D)]} as a prefix.
If we define a predicate that is satisfied when any of the first three elements are not \texttt{(D,D)}, we find that the searchable set does not contain any element satisfying the predicate:
\begin{verbatim}
	> let p xs = xs!!0 /= (D,D) || xs!!1 /= (D,D) || xs!!2 /= (D,D)
	> exists plays p
	==> False
\end{verbatim}
However, by an appropriate choice of predicate we can force the subsequent elements to be anything:
\begin{verbatim}
	> let p' xs = xs!!4 == (D,D)
	> take 6 (search plays p')
	==> [(D,D),(D,D),(D,D),(C,C),(D,D),(C,C)]
\end{verbatim}

The stage in which behaviour changes from \texttt{D} to undetermined is controlled by the precision.
If the precision is reduced to 3 then the searchable set \texttt{plays} changes to the set of streams with \texttt{[(D,D),(D,D)]} as a prefix.
Although increasing the precision above 4 is too slow to test in practice, presumably as the precision tends to infinity the searchable set \texttt{plays} will converge (in some suitable sense) to a singleton set containing only \texttt{[(D,D),(D,D),(D,D),\ldots]}.
This method is thus unable to compute the strategic plays of the (many) other subgame perfect equilibria of IPD.

There are two phenomena that demand an explanation here.
The first is the switch in behaviour, determined by the precision parameter.
This is because we are maximising only up to precision $\frac{1}{N}$: after the switch in behaviour, the difference in payoff caused by different choices is smaller than the error bound.
The reason why \texttt{C} is chosen by default is an implementation detail: ultimately it comes from the ordering \texttt{[C,D]} when we defined \texttt{Move} as an instance of \texttt{Finite}.
If we had instead written \texttt{instance Finite Move where exhaust = [D,C]}, this phenomenon would vanish and we would obtain \texttt{[(D,D),(D,D),\ldots]} by default, but still be able to force \texttt{C} after the (no longer apparent) behaviour switch.

The second phenomenon is far more subtle.
Ignoring rounding errors, it appears that we have obtained a singleton set containing only \texttt{[(D,D),(D,D),\ldots]}.
This is the limit of the solution set of an $n$-stage finitely iterated prisoner's dilemma as $n \to \infty$.
However, the \emph{infinitely} iterated prisoner's dilemma has a much larger solution set.
(This sort of discontinuity is common in game theory.)

We leave the explanation of this as an open problem.
One possible route to a solution is to be clear about the distinction between \emph{subgame perfect equilibria} and \emph{equilibria that can arise by backward induction}.
It is common that game theory texts conflate these two things, or are imprecise about the distinction. 
A possible conjecture is that the set of `backward induction equilibria' is continuous as the number of stages goes to infinity, whereas the set of subgame perfect equilibrium plays has a discontinuity at the limit.
(The former limit is not a known concept in classical game theory, since backward induction for infinite games was only introduced in \cite{escardo10a}.)

The following points all need to be considered for a full understanding of this method:
\begin{itemize}
	\item The game-theoretic meaning of the monad $\J^T_R$ for a \emph{finite} game, where $T$ is the powerset monad
	\item The game-theoretic issues of extending from finite to infinite games, such as discontinuity of the solution set
	\item The topological issues (for correctness) and computability issues (for termination) of using searchable sets
	\item The effect of using $\epsilon$-$\argmax$, which can be seen as reducing an \emph{infinite} game to a \emph{finite but unbounded} one.
\end{itemize}

\bibliographystyle{eptcs}
\bibliography{\string~/Dropbox/Work/refs}

\end{document}